\documentclass[aps,prb,twocolumn]{revtex4-1}
\usepackage{amsmath,amssymb,bm}
\usepackage[dvipdfmx]{graphicx}

\usepackage{txfonts}
\usepackage{amsmath}
\usepackage{bm}
\usepackage{comment}
\usepackage{here}
\usepackage{kantlipsum}
\usepackage{color}
\begin{document}

\title{
Relationship between 
Magnetic Anisotropy Below
Pseudogap Temperature and 
Short-Range Antiferromagnetic Order 
in High-Temperature Cuprate Superconductor
}
\author{Takao Morinari}
\affiliation{
Graduate School of Human and Environmental Studies, Kyoto University, 
Kyoto 606-8501, Japan
}
\date{\today}

\begin{abstract}
The central issue in high-temperature cuprate superconductors
is the pseudogap state appearing below the pseudogap 
temperature $T^*$, which is well above the superconducting
transition temperature.
In this study, we theoretically investigate 
the rapid increase of the magnetic anisotropy
below the pseudogap temperature
detected by the recent torque-magnetometry measurements on 
YBa$_2$Cu$_3$O$_y$ [Y. Sato {\it et al}., 
Nat. Phys., 13, 1074 (2017)].
Applying the spin Green's function formalism including 
the Dzyaloshinskii--Moriya interaction
arising from the buckling of the CuO$_2$ plane,
we obtain results that are in good agreement
with the experiment and find a scaling relationship.
Our analysis suggests that 
the characteristic temperature associated with
the magnetic anisotropy,
which coincides with $T^*$, is
not a phase transition temperature but a crossover temperature
associated with the short-range antiferromagnetic order.
\end{abstract}
\maketitle

       The central issue in high-temperature 
       cuprate superconductors\cite{Keimer2015}
       is the nature and origin of 
       the normal state pseudogap.
       Below the pseudogap temperature, $T^*$,
       which is higher than the superconducting transition temperature,
       $T_c$,
       a partial gap is observed 
       in various experiments.\cite{Timusk1999,Norman2005}
       The key question about the pseudogap is whether
       $T^*$ is a phase transition temperature 
       or a crossover temperature.
       For instance,
       resonant ultrasound spectroscopy measurements 
       exhibited a discontinuous
       change in the temperature dependence of frequency
       supporting that $T^*$ is 
       the phase transition temperature.\cite{Shekhter2013}
       The measurement of the second--harmonic response,
       which detected the inversion symmetry breaking
       below $T^*$,
       also supported the phase transition picture.\cite{Zhao2016}
       Meanwhile, a phenomenological theory 
       describing a crossover scenario
       was proposed,\cite{Yang2006,Rice2012}
       and spectroscopic and
       thermodynamic experiments
       were discussed using a model Green's function
       with doping-dependent parameters.
       On the other hand,
       recent nuclear magnetic resonance\cite{Wu2011,Wu2013}
       and x-ray scattering\cite{Ghiringhelli2012,Achkar2012,Chang2012}
       studies
       reported a symmetry-breaking phase 
       of the charge-density wave order in the pseudogap phase.
       Although the role of this order is unclear,
       it seems to compete with superconductivity\cite{Kharkov2016}
       and it appears at a temperature 
       between $T^*$ and $T_c$.
       It has also been proposed that 
       these orders are intertwined.\cite{Fradkin2015}

       In this Letter, we focus on 
       the recent torque-magnetometry measurements
       on YBa$_2$Cu$_3$O$_y$ (YBCO)
       reporting a rapid increase in
       anisotropic spin susceptibility 
       within the $a-b$ plane
       below $T^*$.\cite{Sato2017}
       A magnetic torque is induced
       if the magnetization ${\bm M}$ of the sample
       is not parallel to the applied magnetic field ${\bm H}$.
       When the magnetic field is rotated
       in the $x-y$ ($a-b$) plane
       by an azimuthal angle $\phi$, 
       the magnetic torque is given by
\begin{eqnarray}
{\tau _\phi } &=& {\mu _0}V{\left( {{\bm{M}} 
			     \times {\bm{H}}} \right)_z} \nonumber \\
 &=& \frac{1}{2}{\mu _0}V{H^2}
  \left[ {\left( {{\chi _{xx}} - {\chi_{yy}}} \right)
   \sin 2\phi  - 2{\chi_{xy}}\cos 2\phi } \right].
\end{eqnarray}
Here, $\mu_0$ is the permeability of vacuum and
$V$ is the sample volume.
The spin susceptibility is denoted by 
${\chi _{\alpha \beta }} 
= \partial {M_\alpha }/\partial \left( {{\mu _0}{H_\beta}} \right)$,
with $\alpha,\beta=x,y$.
For the CuO$_2$ plane with fourfold rotational symmetry, 
C$_4$, we see that $\tau_\phi=0$.
In YBCO, ${\tau _\phi }$ exhibits sinusoidal oscillation
with ${{\chi _{xx}} > {\chi _{yy}}}$ and $\chi_{xy}=0$.\cite{Sato2017}
A rapid increase in the amplitude is observed
below the characteristic temperature 
$T_{\tau}$
that coincides
with the $T^*$ value determined by other experiments.\cite{Sato2017}
The authors in Ref.~\onlinecite{Sato2017} 
conclude that $T_{\tau}$ corresponds to
a nematic phase transition temperature
and thus $T^*$ is also a phase transition temperature.
%

We propose a theory to explain this magnetic torque experiment.
The theory is based on a localized spin model
with anisotropic magnetic interaction.
For this,
we assume the Dzyaloshinskii--Moriya (DM)
interaction\cite{Bonesteel1992,Shekhtman1992,Koshibae1993}
arising from the buckling of the CuO$_2$ plane.
Usually, one may neglect this DM interaction
owing to its energy scale.
However, it breaks the C$_4$ symmetry
and can play an important role for the physical quantities
that do not vanish when the C$_4$ symmetry is broken.
Applying second-order perturbation theory,
we show that $\tau_{\phi}$ is proportional
to cube of the spin susceptibility,
and there is a scaling relationship.
The analysis suggests that 
$T_{\tau}$
is the onset of a short-range antiferromagnetic (AF) order.

In describing the localized spins in the parent 
compound of the cuprate,
the renormalization group analysis of the nonlinear $\sigma$ model
was successful.\cite{Chakravarty1988}
Mean field theories such as Schwinger bosons\cite{Arovas1988}
and modified spin wave theory\cite{Takahashi1989}
also gave a good description of the system.
However, these approaches are useful only
in the low-temperature regime.
At high temperatures around $T^*$, we need to take
a different approach.
Here, we take the spin Green's 
function approach.\cite{Tyablikov1962,Kondo1972,Shimahara1991,Winterfeldt1997,Zavidonov1998,Sadovskii2001}

For the calculation of $\tau_{\phi}$,
we need to compute the following correlation functions:
\begin{eqnarray}
{\left\langle {S_i^xS_j^x} \right\rangle  
- \left\langle {S_i^yS_j^y} \right\rangle }
&=& {{\mathop{\rm Re}\nolimits} 
\left\langle {S_i^ + S_j^ + } \right\rangle }, \\
{\left\langle {S_i^xS_j^y} \right\rangle  
+ \left\langle {S_i^yS_j^x} \right\rangle }
&=& {{\mathop{\rm Im}\nolimits} 
\left\langle {S_i^ + S_j^ + } \right\rangle }.
\end{eqnarray}
Here, $S_j^{\alpha}$ ($\alpha=x,y$) denotes the $\alpha$ component 
of the spin moment at site $j$.
Note that these correlation functions depend on $i-j$
because of the translational invariance in the pseudogap phase.
In the absence of any magnetically anisotropic term,
the right-hand sides of these equations vanish.
The Hamiltonian for the localized $S=1/2$ moments, 
on inclusion of the DM interaction mentioned above, is given by
\begin{equation}
 {\mathcal H} = {J_{p}}\sum\limits_{
  \left\langle {i,j}
  \right\rangle } {{{\bm{S}}_i} \cdot {{\bm{S}}_j}}  
  + \sum\limits_{\left\langle {i,j} \right\rangle }
  {{{\bm{D}}_{ij}} \cdot \left( {{{\bm{S}}_i} \times {{\bm{S}}_j}}
  \right)}.
\end{equation}
Here, $J_p$ is the exchange interaction 
between nearest-neighbor spins,
which is assumed to depend on the doped hole concentration, $p$.
The three-dimensional vector 
${\bm{D}}_{ij} = {\bm{D}}_{i-j}$ is the DM vector on the bond connecting
sites $i$ and $j$.
For the case of $D_{i - j}^z = 0$,
the DM interaction term is rewritten as
\begin{equation}
{\mathcal H}_{{\rm{DM}}} = \sum\limits_i 
\sum\limits_{\delta  = \widehat a,\widehat b}
{\left( {{g_\delta }S_i^ - S_{i + \delta }^z + H.c.} \right)}.
\label{eq_H_DM}
\end{equation}
Here, $\widehat a$ and 
$\widehat b$ are the displacement vectors
along the $a$ and $b$ axes, respectively,
and
${g_\delta } = \left( {iD_\delta ^{x} - D_\delta^{y}} \right)/2$,
with $D_\delta^{\alpha}$ being 
the $\alpha$ component of the DM vector.
It is obvious from Eq.~(\ref{eq_H_DM}),
that its first-order contribution to 
$\left\langle {S_i^ + S_j^ + } \right\rangle$ vanishes,
but the second-order contribution does not.

Now, we define the following Matsubara Green's function:
\begin{equation}
 {G_{i-j}}\left( \tau  \right) =  
  - \left\langle {{T_\tau }S_i^ + \left(\tau  \right)
     S_j^ - \left( 0 \right)} \right\rangle,
\end{equation}
with $\tau$ being the imaginary time.
Taking the derivative of $G_{i-j}(\tau)$ with respect to $\tau$ twice,
and then applying the Tyablikov approximation
and the Fourier transform,
we obtain\cite{Kondo1972,Shimahara1991}
\begin{equation}
 G_{\bm{k}}\left( {i{\omega _n}} \right) 
  = \frac{ 4J_p {c_1}
  \left( 1 - {\gamma _{\bm {k}}} \right) }
  { \left( {i{\omega _n}} \right)^2 -
  \omega _{\bm{k}}^2},
\end{equation}
with $\omega_n$ denoting the Matsubara frequency and
\begin{equation}
{c_{i-j}} = 4\left\langle {S_i^zS_j^z} \right\rangle  = 2\left\langle
{S_i^ + S_j^ - } \right\rangle  = 2\left\langle {S_i^ - S_j^ + }
\right\rangle.
\end{equation}
(Hereafter, we set $\hbar=1$ and the lattice constant is set to unity.)
The spin excitation energy ${\omega _{\bm{k}}}$
is given by
\begin{equation}
{\omega _{\bm{k}}} = \sqrt {8\alpha \left| {{c_1}} \right|}
 {J_{p}}\sqrt {\left( {1 - {\gamma _{\bm{k}}}} \right)\left( {1 +
  \Delta  + {\gamma _{\bm{k}}}} \right)},
\end{equation}
with ${\gamma _{\bm{k}}} = \left( {\cos {k_x} + \cos {k_y}} \right)/2$
and 
$\Delta  = \left( {1 - \alpha {c_1} + 3\alpha {c'_2}} \right)/\left(
  {4\alpha \left| {{c_1}} \right|} \right) - 1$.
The parameter $\alpha$ is introduced
while applying the Tyablikov approximation,\cite{Kondo1972}
which is interpreted as a vertex correction.\cite{Shimahara1991}
The parameter ${c'_2}$ is defined by
${c'_2} = \sum\limits_{\delta '\left( { \ne  - \delta } \right)}
  {{c_{\delta  + \delta '}}} /3$.
The parameters $c_1$, $\alpha$, and ${c'_2}$
are determined by solving 
the following self-consistent equations\cite{Shimahara1991}:
\begin{equation}
1 =  - \frac{{4{J_p}{c_1}}}{N}\sum\limits_{\bm{k}} {\frac{{1 - {\gamma
 _{\bm{k}}}}}{{{\omega _{\bm{k}}}}}
\coth \left( {\frac{{{\omega _{\bm{k}}}}}{2k_{B} T}}
				  \right)},
\end{equation}
\begin{equation}
{c_1} =  - \frac{{4{J_p}{c_1}}}{N}\sum\limits_{\bm{k}} {\frac{{{\gamma
 _{\bm{k}}}\left( {1 - {\gamma _{\bm{k}}}} \right)}}
{{{\omega _{\bm{k}}}}}\coth
 \left( {\frac{{{\omega _{\bm{k}}}}}{2k_{B} T}} \right)},
\end{equation}
\begin{equation}
 \frac{{3{c'_2} + 1}}{4} =  -
  \frac{{4{J_p}{c_1}}}{N}\sum\limits_{\bm{k}} {\frac{{\gamma
  _{\bm{k}}^2\left( {1 - {\gamma _{\bm{k}}}} \right)}}{{{\omega
  _{\bm{k}}}}}\coth \left( {\frac{{{\omega _{\bm{k}}}}}{2k_{B} T}} 
\right)}.
\end{equation}
Here, $N$ is the number of the lattice sites,
and $k_{B}$ is the Boltzmann constant.

The second-order perturbative calculation
with respect to $H_{DM}$ gives
\begin{eqnarray}
 \left\langle {S_i^ + S_j^ + } \right\rangle  
  &=&
  \frac{{{k_{\rm{B}}}T}}{{2N}}\sum\limits_{\bm{k}} 
  {\sum\limits_{\delta,\delta '} 
  {{g_\delta }{g_{\delta '}}{e^{i{\bm{k}} \cdot 
  \left( {\delta - \delta '} \right)}}}
  \sum\limits_{i{\omega _n}} {{e^{i{\bm{k}} \cdot 
  \left( {{{\bm{R}}_i} - {{\bm{R}}_j}} \right)}}} } \nonumber \\
 & & \times {G_{\bm{k}}}\left( {i{\omega _n}} \right)
  {G_{ - {\bm{k}}}}\left( { - i{\omega _n}} \right)
  {G_{ - {\bm{k}}}}\left( {i{\omega _n}} \right),
\end{eqnarray}
where $\bm{R}_i$ denotes the position of site $i$.
The summation over $i$ shows that we need only
the ${\bm{k}}=0$ term.
The terms with $\omega_n \neq 0$ vanish if we set 
${\bm{k}}=0$.
Therefore, we may set $\omega_n=0$, and then ${\bm{k}}=0$.
The result is
\begin{equation}
 \frac{1}{N} \sum\limits_i {\left\langle {S_i^ + S_j^ + } \right\rangle }  
= \frac{{{k_{\rm{B}}}T}}{{16J_{p}^3{\alpha ^3}
{{\left( {2 + \Delta }\right)}^3}}}
\Gamma,
\end{equation}
with
$\Gamma  = {\left( {{g_{\widehat a}} + {g_{\widehat b}}} \right)^2}$.
By using this result, we obtain
\begin{equation}
\Delta {\chi} \equiv \frac{{{\tau _\phi }}}{{{\mu
  _0}V{H^2}/2}} = \frac{{{\mu _0}\mu _B^2}}{{2{v_c}}}\frac{{{\Gamma
  _\parallel }\sin 2\phi  - {\Gamma _ \bot }\cos 2\phi }}{{J_p^3{\alpha
  ^3}{{\left( {2 + \Delta } \right)}^3}}},
\label{eq_chi_parallel}
\end{equation}
where ${v_c}$ is the unit cell volume per CuO$_2$ plane,
and
${\Gamma _\parallel } = {\mathop{\rm Re}\nolimits} \Gamma$
and
${\Gamma _ \bot } = {\mathop{\rm Im}\nolimits} \Gamma$.
$\Delta {\chi}$ oscillates with two components:
one is proportional to $\sin 2\phi$,
and the other is proportional to $\cos 2\phi$.
We note that\cite{Shimahara1991}
\begin{equation}
 \frac{1}{N}\sum\limits_i 
  {\left\langle {S_i^ + S_j^ - } \right\rangle }  
  = \frac{1}{{2{J_p}\alpha \left( {2 + \Delta } \right)}}.
\end{equation}
Therefore, the right-hand side of Eq.~(\ref{eq_chi_parallel})
is proportional to the cube of the spin susceptibility.

\begin{figure}[htbp]
  \centering
 \includegraphics[width=0.4\textwidth]{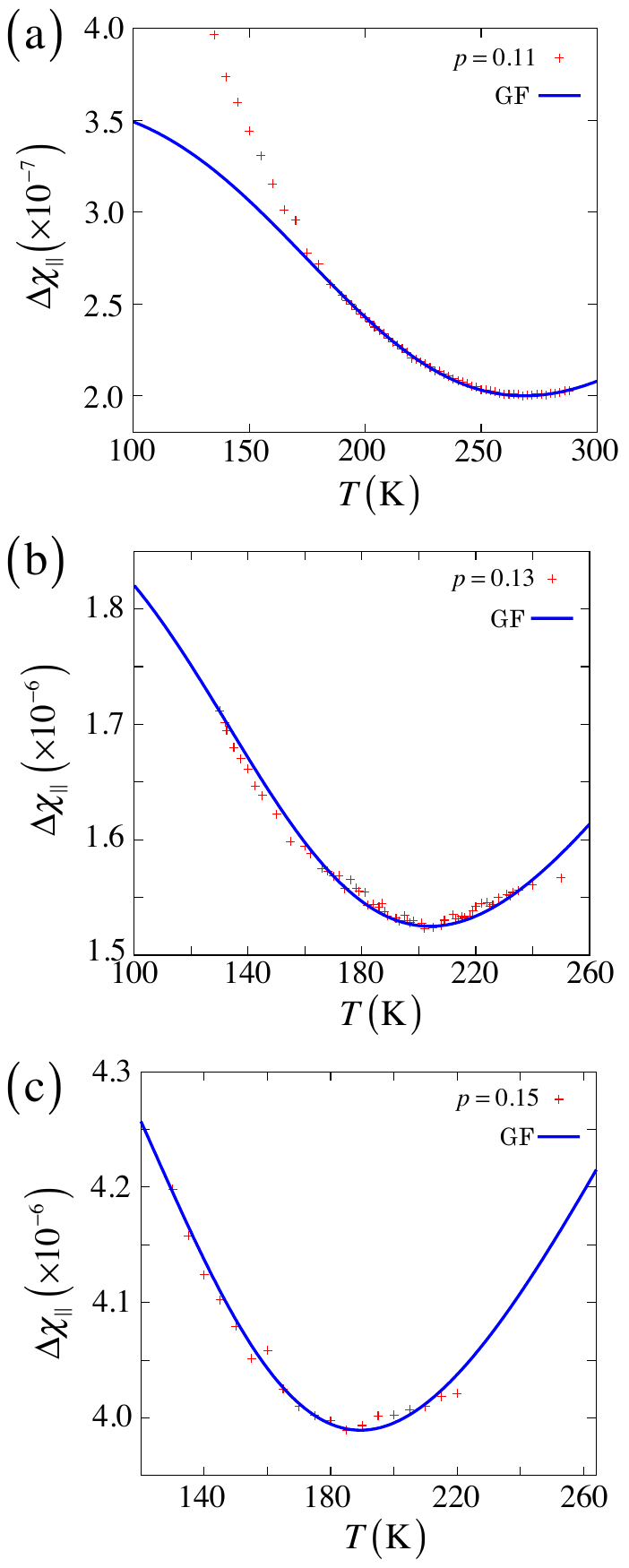}
  \caption{(color online)
 Comparisons between the formula (\ref{eq_chi_parallel})
 and the experiments\cite{Sato2017}
 for hole concentrations (a) $p=0.11$, (b) $p=0.13$,
 and (c) $p=0.15$.
 The solid lines represent the theory based on the spin Green's function.
 }
  \label{fig:each}
\end{figure}

Now we apply the theory to the experiment.\cite{Sato2017}
For YBCO, 
$D_{\widehat a}^{y}\ne 0$,
$D_{\widehat b}^{x}\ne 0$,
and the other components are negligible.\cite{Bonesteel1993}
Thus, ${\Gamma _\parallel } \neq 0$
and 
${\Gamma _ \bot } =  
- \left[ {D_a^{\left( x \right)}D_a^{\left( y \right)} +
D_b^{\left( x \right)}D_b^{\left( y \right)}} \right]/2 = 0$.
Therefore, we find $\tau_{\phi} \propto \sin 2\phi$,
which is the oscillation pattern observed in the
experiment.\cite{Sato2017}
Hereafter, we consider the case $\Gamma_\bot = 0$,
and denote $\Delta \chi$ as $\Delta \chi_\parallel$.
The theoretical formula (\ref{eq_chi_parallel}) is compared with
the experiment\cite{Sato2017}
with the fitting parameters $J_p$ and ${\Gamma _\parallel }$
by including a constant term consisting of a temperature-independent
paramagnetic component.
The results shown in Fig.~\ref{fig:each}
demonstrate that 
the theory is in good agreement with the experiment.
From the fitting, we found
$J_{0.11}=241$ K,
$J_{0.13}=183$ K,
and 
$J_{0.15}=170$ K
as the values of $J_p$ for $p$ = 0.11, 0.13, and 0.15 respectively.
The value of $J_p$ decreases as $p$ is increased.
This monotonic change in $J_p$ as a function of $p$ 
was also suggested 
from an analysis of the spin susceptibility 
and a scaling was found
in La$_{2-x}$Sr$_x$CuO$_{4-y}$.\cite{Johnston1989,Nakano1994}
For $p=0.11$, there is a discrepancy between theory
and the experiment at low temperatures.
This is because the spin Green's function approach
is not reliable at low temperatures.\cite{Shimahara1991}
We note that this discrepancy starts from $0.40 J_p$ below
the minimum of $\Delta \chi_\parallel$.
The data for $p=0.13$ and $p=0.15$ are well above this value.

\begin{figure}[htbp]
  \centering
\includegraphics[width=0.4\textwidth]{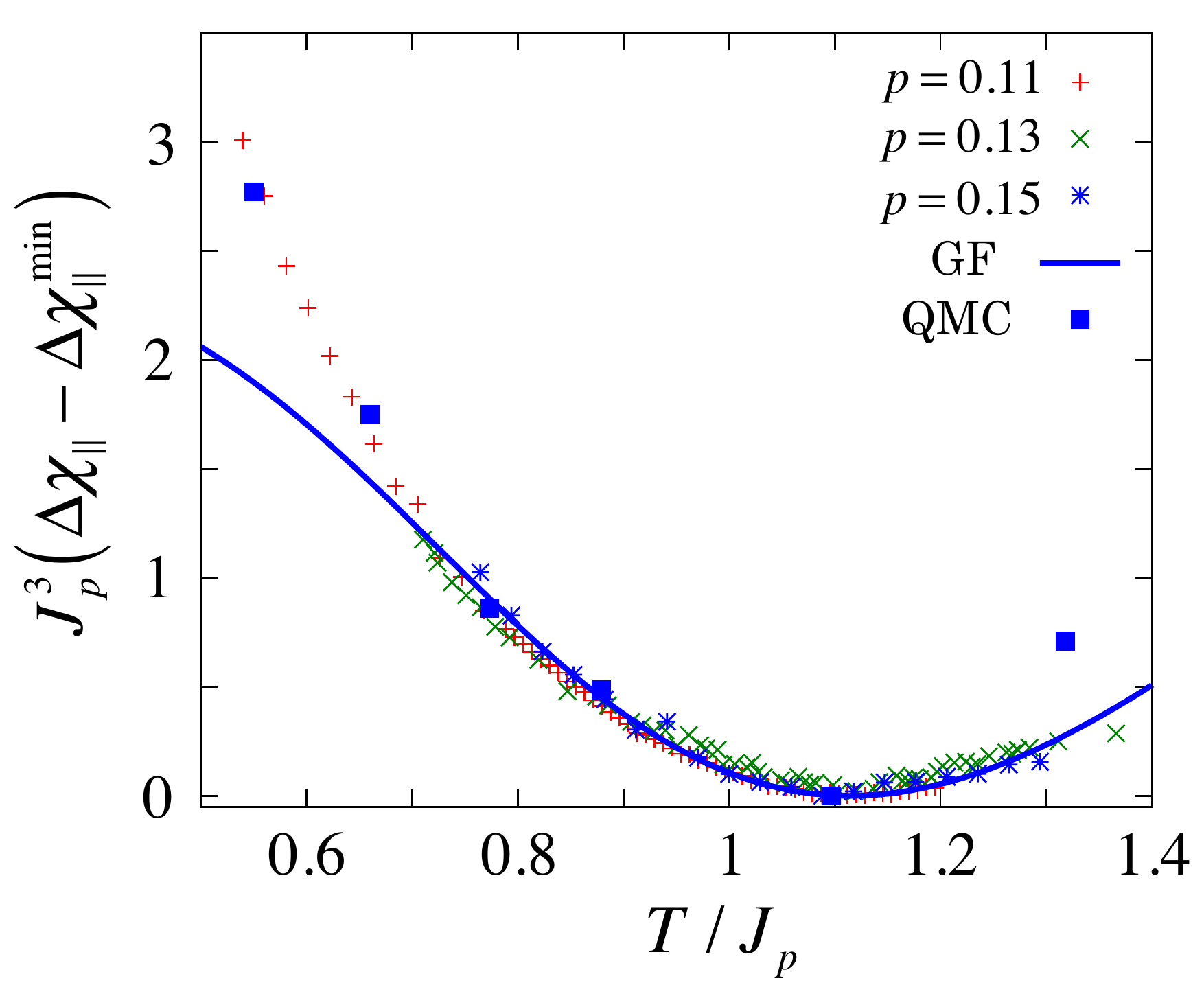}
  \caption{(color online) 
 Scaling relationship suggested from the formula 
 (\ref{eq_chi_parallel}).
 $\Delta \chi _\parallel ^{\min }$
 is the minimum of $\Delta \chi _\parallel$
 in Fig.~\ref{fig:each}.
 The unit of the vertical axis is K$^3$.
 For the values of $J_p$, we take 
 $J_{0.11}=241$ K,
 $J_{0.13}=183$ K,
 and
 $J_{0.15}=170$ K
 for the experimental data.
 The values computed by using QMC result\cite{Okabe1988} are also shown.
 }
  \label{fig:normalized}
\end{figure}

From the formula (\ref{eq_chi_parallel}),
we see that
$J_p^3\Delta {\chi _\parallel }$
is independent of $J_p$.
In order to remove constant components coming from doped holes,
we subtract its minimum value from $\Delta {\chi _\parallel }$,
and then plot it
as a function of the normalized temperature
in Fig.~\ref{fig:normalized}.
All the experimental data fall on a single curve.
From this analysis, we may conclude that
$T_{\tau} \simeq 1.1 J_p$.
This characteristic temperature has a simple interpretation.
The AF correlation length 
of the AF Heisenberg model with the exchange interaction $J_p$
is given by\cite{Takahashi1989}
\begin{equation}
 {\xi _{{\rm{AF}}}/a} \simeq
  \frac{{0.819}}{{T/{J_p}}}\exp \left( {\frac{{1.10}}{{T/{J_p}}}} \right),
\end{equation}
where $a$ is the lattice constant.
From this formula, we find 
$\xi_{\rm AF} \simeq 2 a$ at $T=T_{\tau}$.
In Fig.~\ref{fig:normalized}
we also plot the values computed using quantum Monte Carlo (QMC)
results for the uniform spin susceptibility $\chi$ on the
square lattice AF Heisenberg model.\cite{Okabe1988}
These values are in good agreement with
the data of $p=0.11$ at low temperatures.
However, the point computed from the QMC data around $T/J_p = 1.3$ 
does not agree with the experiment and the Green's function result.
We note that we find 
${\Gamma _\parallel } < 0$
from the fact that the magnitude of the DM vector
is proportional 
to the difference in the lattice constants in the orthorhombic
phase of YBCO.
This is consistent with the experiment
because the maximum of $\chi$ corresponds to
the minimum of $\Delta {\chi _\parallel }$.
We also note that the experimental data
seem to be convex upward for $T>T_{\tau}$
at $p=0.13$ and $p=0.15$.
However, a similar behavior is not discernible for $p=0.11$.
It might be related to the effect of doped holes
and/or CuO chains.

Now we discuss the value of $\Gamma_\parallel$.
From the analysis shown in Fig.~\ref{fig:normalized}, 
we find $\sqrt{|\Gamma_\parallel|} \simeq 100$ K.
This apparently is too large 
if $\Gamma_\parallel$ is associated with
the buckling of the CuO$_2$ plane.
Here, we need to include the effect of the doped holes.
The exchange coupling between doped hole spins
and the localized spins is described by
${\mathcal H}_K = {J_K}\sum\limits_j {{{\bm{S}}_j} 
\cdot \left( {c_j^\dag {\bm \sigma} {c_j}} \right)}$,
where 
${J_K} = t_{dp}^2/\left( {{U_d} - \Delta } \right) + t_{dp}^2/\left(
{{U_p} + \Delta } \right)$,
where $t_{dp}$ is the nearest-neighbor Cu-O hopping
and
$U_d$ ($U_p$) is the Cu(O)-site Coulomb repulsion.\cite{ZhangRice1988,Tohyama1990,Matsukawa1989}
$\Delta$ is the energy difference
between the O-site energy and the Cu-site energy.
The two-component operator $c_j^{\dagger}$ ($c_j$) is 
the creation (annihilation) operator 
of the doped hole at site $j$,
and ${\bm \sigma}$ is the three component vector
of the Pauli matrices.
The easiest way to include ${\mathcal H}_K$
is the coherent state path integral.
By integrating out the doped hole fields,
we find that the spin susceptibility $\chi$
is enhanced as
$\chi /\left( {1 - \eta \chi } \right)$
with $\eta  = 3J_K^2\chi_0^{\rm{h}}$.
Here, $\chi_0^{\rm{h}}$
is the uniform spin susceptibility of the doped holes.
Unfortunately no reliable theoretical formula
for $\chi_0^{\rm{h}}$ is available.
Therefore, we use the formula for the non-interacting system,
which is proportional to the density of states,
and approximate it
as $\chi_0^{\rm{h}} \sim 1/t$
where $t \sim t_{dp}^2/\Delta$ 
is the effective hopping parameter 
of the doped holes.\cite{ZhangRice1988,Tohyama1990,Matsukawa1989}
Using the parameter values evaluated 
for the CuO$_2$ plane,\cite{Tohyama1990,Eskes1989,Hybertsen1990,Matsukawa1989}
we find that $\eta/J_p \sim 10$.
With this value of $\eta/J_p$,
$\sqrt{|\Gamma_\parallel|} \sim 2$ K.
For $\eta/J_p = 9$, 
$\sqrt{|\Gamma_\parallel|} \sim 15$ K.
Although this is an approximate estimate,
these values appear to be reasonable
from the fact that $\Gamma_\parallel$
is proportional to the difference between the lattice constants
along the $a$ and $b$ axes
and also the buckling angle.

To conclude, 
we have shown that 
the result of the theory based on 
the spin Green's function 
with the DM interaction 
is in good agreement with
the recent torque-magnetometry measurements 
of YBCO.\cite{Sato2017}
There is a clear scaling relationship 
as shown in Fig.~\ref{fig:normalized}.
Our analysis shows that the magnetic anisotropy
increases rapidly below $T_{\tau} \simeq 1.1 J_p$ 
at which $\xi_{\rm AF} \simeq 2 a$.
Therefore, $T_{\tau}$ is a crossover temperature associated
with the short-range AF order, in contrast to the claim 
in Ref.~\onlinecite{Sato2017} where
$T_{\tau}$ is interpreted as an onset of a nematic phase transition. 
Given the experimental fact that $T_{\tau}$ 
coincides with the onset temperature
of the pseudogap, the pseudogap may also be a crossover phenomenon.

\section*{Acknowledgments}
The author thanks Y. Matsuda and Y. Sato
for providing the experimental data.

\bibliographystyle{apsrev4-1}
\bibliography{../../refs/tm_lib201801}

\begin{thebibliography}{36}%
\makeatletter
\providecommand \@ifxundefined [1]{%
 \@ifx{#1\undefined}
}%
\providecommand \@ifnum [1]{%
 \ifnum #1\expandafter \@firstoftwo
 \else \expandafter \@secondoftwo
 \fi
}%
\providecommand \@ifx [1]{%
 \ifx #1\expandafter \@firstoftwo
 \else \expandafter \@secondoftwo
 \fi
}%
\providecommand \natexlab [1]{#1}%
\providecommand \enquote  [1]{``#1''}%
\providecommand \bibnamefont  [1]{#1}%
\providecommand \bibfnamefont [1]{#1}%
\providecommand \citenamefont [1]{#1}%
\providecommand \href@noop [0]{\@secondoftwo}%
\providecommand \href [0]{\begingroup \@sanitize@url \@href}%
\providecommand \@href[1]{\@@startlink{#1}\@@href}%
\providecommand \@@href[1]{\endgroup#1\@@endlink}%
\providecommand \@sanitize@url [0]{\catcode `\\12\catcode `\$12\catcode
  `\&12\catcode `\#12\catcode `\^12\catcode `\_12\catcode `\%12\relax}%
\providecommand \@@startlink[1]{}%
\providecommand \@@endlink[0]{}%
\providecommand \url  [0]{\begingroup\@sanitize@url \@url }%
\providecommand \@url [1]{\endgroup\@href {#1}{\urlprefix }}%
\providecommand \urlprefix  [0]{URL }%
\providecommand \Eprint [0]{\href }%
\@ifxundefined \urlstyle {%
  \providecommand \doi  [0]{\begingroup \@sanitize@url \@doi}%
  \providecommand \@doi [1]{\endgroup \@@startlink {\doibase
  #1}doi:\discretionary {}{}{}#1\@@endlink }%
}{%
  \providecommand \doi  [0]{doi:\discretionary{}{}{}\begingroup
  \urlstyle{rm}\Url }%
}%
\providecommand \doibase [0]{http://dx.doi.org/}%
\providecommand \Doi [0]{\begingroup \@sanitize@url \@Doi }%
\providecommand \@Doi  [1]{\endgroup\@@startlink{\doibase#1}\@@Doi}%
\providecommand \@@Doi [1]{#1\@@endlink}%
\providecommand \selectlanguage [0]{\@gobble}%
\providecommand \bibinfo  [0]{\@secondoftwo}%
\providecommand \bibfield  [0]{\@secondoftwo}%
\providecommand \translation [1]{[#1]}%
\providecommand \BibitemOpen [0]{}%
\providecommand \bibitemStop [0]{}%
\providecommand \bibitemNoStop [0]{.\EOS\space}%
\providecommand \EOS [0]{\spacefactor3000\relax}%
\providecommand \BibitemShut  [1]{\csname bibitem#1\endcsname}%
\bibitem [{\citenamefont {Keimer}\ \emph {et~al.}(2015)\citenamefont {Keimer},
  \citenamefont {Kivelson}, \citenamefont {Norman}, \citenamefont {Uchida},\
  and\ \citenamefont {Zaanen}}]{Keimer2015}%
  \BibitemOpen
  \bibfield  {author} {\bibinfo {author} {\bibfnamefont {B.}~\bibnamefont
  {Keimer}}, \bibinfo {author} {\bibfnamefont {S.}~\bibnamefont {Kivelson}},
  \bibinfo {author} {\bibfnamefont {M.}~\bibnamefont {Norman}}, \bibinfo
  {author} {\bibfnamefont {S.}~\bibnamefont {Uchida}}, \ and\ \bibinfo {author}
  {\bibfnamefont {J.}~\bibnamefont {Zaanen}},\ }\href@noop {} {\bibfield
  {journal} {\bibinfo  {journal} {Nature},\ }\textbf {\bibinfo {volume}
  {518}},\ \bibinfo {pages} {179} (\bibinfo {year} {2015})}\BibitemShut
  {NoStop}%
\bibitem [{\citenamefont {Timusk}\ and\ \citenamefont
  {Statt}(1999)}]{Timusk1999}%
  \BibitemOpen
  \bibfield  {author} {\bibinfo {author} {\bibfnamefont {T.}~\bibnamefont
  {Timusk}}\ and\ \bibinfo {author} {\bibfnamefont {B.~W.}\ \bibnamefont
  {Statt}},\ }\href@noop {} {\bibfield  {journal} {\bibinfo  {journal} {Rep.
  Prog. Phys.},\ }\textbf {\bibinfo {volume} {62}},\ \bibinfo {pages} {61}
  (\bibinfo {year} {1999})}\BibitemShut {NoStop}%
\bibitem [{\citenamefont {Norman}\ \emph {et~al.}(2005)\citenamefont {Norman},
  \citenamefont {Pines},\ and\ \citenamefont {Kallin}}]{Norman2005}%
  \BibitemOpen
  \bibfield  {author} {\bibinfo {author} {\bibfnamefont {M.~R.}\ \bibnamefont
  {Norman}}, \bibinfo {author} {\bibfnamefont {D.}~\bibnamefont {Pines}}, \
  and\ \bibinfo {author} {\bibfnamefont {C.}~\bibnamefont {Kallin}},\ }\Doi
  {10.1080/00018730500459906} {\bibfield  {journal} {\bibinfo  {journal} {Adv.
  Phys.},\ }\textbf {\bibinfo {volume} {54}},\ \bibinfo {pages} {715} (\bibinfo
  {year} {2005})}\BibitemShut {NoStop}%
\bibitem [{\citenamefont {Shekhter}\ \emph {et~al.}(2013)\citenamefont
  {Shekhter}, \citenamefont {Ramshaw}, \citenamefont {Liang}, \citenamefont
  {Hardy}, \citenamefont {Bonn}, \citenamefont {Balakirev}, \citenamefont
  {McDonald}, \citenamefont {Betts}, \citenamefont {Riggs},\ and\ \citenamefont
  {Migliori}}]{Shekhter2013}%
  \BibitemOpen
  \bibfield  {author} {\bibinfo {author} {\bibfnamefont {A.}~\bibnamefont
  {Shekhter}}, \bibinfo {author} {\bibfnamefont {B.~J.}\ \bibnamefont
  {Ramshaw}}, \bibinfo {author} {\bibfnamefont {R.}~\bibnamefont {Liang}},
  \bibinfo {author} {\bibfnamefont {W.~N.}\ \bibnamefont {Hardy}}, \bibinfo
  {author} {\bibfnamefont {D.~A.}\ \bibnamefont {Bonn}}, \bibinfo {author}
  {\bibfnamefont {F.~F.}\ \bibnamefont {Balakirev}}, \bibinfo {author}
  {\bibfnamefont {R.~D.}\ \bibnamefont {McDonald}}, \bibinfo {author}
  {\bibfnamefont {J.~B.}\ \bibnamefont {Betts}}, \bibinfo {author}
  {\bibfnamefont {S.~C.}\ \bibnamefont {Riggs}}, \ and\ \bibinfo {author}
  {\bibfnamefont {A.}~\bibnamefont {Migliori}},\ }\Doi {10.1038/nature12165}
  {\bibfield  {journal} {\bibinfo  {journal} {Nature},\ }\textbf {\bibinfo
  {volume} {498}},\ \bibinfo {pages} {75} (\bibinfo {year} {2013})}\BibitemShut
  {NoStop}%
\bibitem [{\citenamefont {Zhao}\ \emph {et~al.}(2016)\citenamefont {Zhao},
  \citenamefont {Belvin}, \citenamefont {Liang}, \citenamefont {Bonn},
  \citenamefont {Hardy}, \citenamefont {Armitage},\ and\ \citenamefont
  {Hsieh}}]{Zhao2016}%
  \BibitemOpen
  \bibfield  {author} {\bibinfo {author} {\bibfnamefont {L.}~\bibnamefont
  {Zhao}}, \bibinfo {author} {\bibfnamefont {C.~A.}\ \bibnamefont {Belvin}},
  \bibinfo {author} {\bibfnamefont {R.}~\bibnamefont {Liang}}, \bibinfo
  {author} {\bibfnamefont {D.~A.}\ \bibnamefont {Bonn}}, \bibinfo {author}
  {\bibfnamefont {W.~N.}\ \bibnamefont {Hardy}}, \bibinfo {author}
  {\bibfnamefont {N.~P.}\ \bibnamefont {Armitage}}, \ and\ \bibinfo {author}
  {\bibfnamefont {D.}~\bibnamefont {Hsieh}},\ }\Doi {10.1038/nphys3962}
  {\bibfield  {journal} {\bibinfo  {journal} {Nat. Phys.},\ }\textbf {\bibinfo
  {volume} {13}},\ \bibinfo {pages} {250} (\bibinfo {year} {2016})}\BibitemShut
  {NoStop}%
\bibitem [{\citenamefont {Yang}\ \emph {et~al.}(2006)\citenamefont {Yang},
  \citenamefont {Rice},\ and\ \citenamefont {Zhang}}]{Yang2006}%
  \BibitemOpen
  \bibfield  {author} {\bibinfo {author} {\bibfnamefont {K.-Y.}\ \bibnamefont
  {Yang}}, \bibinfo {author} {\bibfnamefont {T.~M.}\ \bibnamefont {Rice}}, \
  and\ \bibinfo {author} {\bibfnamefont {F.-C.}\ \bibnamefont {Zhang}},\ }\Doi
  {10.1103/physrevb.73.174501} {\bibfield  {journal} {\bibinfo  {journal}
  {Phys. Rev. B},\ }\textbf {\bibinfo {volume} {73}},\ \bibinfo {pages}
  {174501} (\bibinfo {year} {2006})}\BibitemShut {NoStop}%
\bibitem [{\citenamefont {Rice}\ \emph {et~al.}(2012)\citenamefont {Rice},
  \citenamefont {Yang},\ and\ \citenamefont {Zhang}}]{Rice2012}%
  \BibitemOpen
  \bibfield  {author} {\bibinfo {author} {\bibfnamefont {T.~M.}\ \bibnamefont
  {Rice}}, \bibinfo {author} {\bibfnamefont {K.-Y.}\ \bibnamefont {Yang}}, \
  and\ \bibinfo {author} {\bibfnamefont {F.~C.}\ \bibnamefont {Zhang}},\ }\Doi
  {10.1088/0034-4885/75/1/016502} {\bibfield  {journal} {\bibinfo  {journal}
  {Rep. Prog. Phys.},\ }\textbf {\bibinfo {volume} {75}},\ \bibinfo {pages}
  {016502} (\bibinfo {year} {2012})}\BibitemShut {NoStop}%
\bibitem [{\citenamefont {Wu}\ \emph {et~al.}(2011)\citenamefont {Wu},
  \citenamefont {Mayaffre}, \citenamefont {Krämer}, \citenamefont
  {Horvati{\'{c}}}, \citenamefont {Berthier}, \citenamefont {Hardy},
  \citenamefont {Liang}, \citenamefont {Bonn},\ and\ \citenamefont
  {Julien}}]{Wu2011}%
  \BibitemOpen
  \bibfield  {author} {\bibinfo {author} {\bibfnamefont {T.}~\bibnamefont
  {Wu}}, \bibinfo {author} {\bibfnamefont {H.}~\bibnamefont {Mayaffre}},
  \bibinfo {author} {\bibfnamefont {S.}~\bibnamefont {Krämer}}, \bibinfo
  {author} {\bibfnamefont {M.}~\bibnamefont {Horvati{\'{c}}}}, \bibinfo
  {author} {\bibfnamefont {C.}~\bibnamefont {Berthier}}, \bibinfo {author}
  {\bibfnamefont {W.~N.}\ \bibnamefont {Hardy}}, \bibinfo {author}
  {\bibfnamefont {R.}~\bibnamefont {Liang}}, \bibinfo {author} {\bibfnamefont
  {D.~A.}\ \bibnamefont {Bonn}}, \ and\ \bibinfo {author} {\bibfnamefont
  {M.-H.}\ \bibnamefont {Julien}},\ }\Doi {10.1038/nature10345} {\bibfield
  {journal} {\bibinfo  {journal} {Nature},\ }\textbf {\bibinfo {volume}
  {477}},\ \bibinfo {pages} {191} (\bibinfo {year} {2011})}\BibitemShut
  {NoStop}%
\bibitem [{\citenamefont {Wu}\ \emph {et~al.}(2013)\citenamefont {Wu},
  \citenamefont {Mayaffre}, \citenamefont {Krämer}, \citenamefont
  {Horvati{\'{c}}}, \citenamefont {Berthier}, \citenamefont {Kuhns},
  \citenamefont {Reyes}, \citenamefont {Liang}, \citenamefont {Hardy},
  \citenamefont {Bonn},\ and\ \citenamefont {Julien}}]{Wu2013}%
  \BibitemOpen
  \bibfield  {author} {\bibinfo {author} {\bibfnamefont {T.}~\bibnamefont
  {Wu}}, \bibinfo {author} {\bibfnamefont {H.}~\bibnamefont {Mayaffre}},
  \bibinfo {author} {\bibfnamefont {S.}~\bibnamefont {Krämer}}, \bibinfo
  {author} {\bibfnamefont {M.}~\bibnamefont {Horvati{\'{c}}}}, \bibinfo
  {author} {\bibfnamefont {C.}~\bibnamefont {Berthier}}, \bibinfo {author}
  {\bibfnamefont {P.~L.}\ \bibnamefont {Kuhns}}, \bibinfo {author}
  {\bibfnamefont {A.~P.}\ \bibnamefont {Reyes}}, \bibinfo {author}
  {\bibfnamefont {R.}~\bibnamefont {Liang}}, \bibinfo {author} {\bibfnamefont
  {W.~N.}\ \bibnamefont {Hardy}}, \bibinfo {author} {\bibfnamefont {D.~A.}\
  \bibnamefont {Bonn}}, \ and\ \bibinfo {author} {\bibfnamefont {M.-H.}\
  \bibnamefont {Julien}},\ }\Doi {10.1038/ncomms3113} {\bibfield  {journal}
  {\bibinfo  {journal} {Nat. Commun.},\ }\textbf {\bibinfo {volume} {4}}
  (\bibinfo {year} {2013})},\ \doi {10.1038/ncomms3113}\BibitemShut {NoStop}%
\bibitem [{\citenamefont {Ghiringhelli}\ \emph {et~al.}(2012)\citenamefont
  {Ghiringhelli}, \citenamefont {Tacon}, \citenamefont {Minola}, \citenamefont
  {Blanco-Canosa}, \citenamefont {Mazzoli}, \citenamefont {Brookes},
  \citenamefont {Luca}, \citenamefont {Frano}, \citenamefont {Hawthorn},
  \citenamefont {He}, \citenamefont {Loew}, \citenamefont {Sala}, \citenamefont
  {Peets}, \citenamefont {Salluzzo}, \citenamefont {Schierle}, \citenamefont
  {Sutarto}, \citenamefont {Sawatzky}, \citenamefont {Weschke}, \citenamefont
  {Keimer},\ and\ \citenamefont {Braicovich}}]{Ghiringhelli2012}%
  \BibitemOpen
  \bibfield  {author} {\bibinfo {author} {\bibfnamefont {G.}~\bibnamefont
  {Ghiringhelli}}, \bibinfo {author} {\bibfnamefont {M.~L.}\ \bibnamefont
  {Tacon}}, \bibinfo {author} {\bibfnamefont {M.}~\bibnamefont {Minola}},
  \bibinfo {author} {\bibfnamefont {S.}~\bibnamefont {Blanco-Canosa}}, \bibinfo
  {author} {\bibfnamefont {C.}~\bibnamefont {Mazzoli}}, \bibinfo {author}
  {\bibfnamefont {N.~B.}\ \bibnamefont {Brookes}}, \bibinfo {author}
  {\bibfnamefont {G.~M.~D.}\ \bibnamefont {Luca}}, \bibinfo {author}
  {\bibfnamefont {A.}~\bibnamefont {Frano}}, \bibinfo {author} {\bibfnamefont
  {D.~G.}\ \bibnamefont {Hawthorn}}, \bibinfo {author} {\bibfnamefont
  {F.}~\bibnamefont {He}}, \bibinfo {author} {\bibfnamefont {T.}~\bibnamefont
  {Loew}}, \bibinfo {author} {\bibfnamefont {M.~M.}\ \bibnamefont {Sala}},
  \bibinfo {author} {\bibfnamefont {D.~C.}\ \bibnamefont {Peets}}, \bibinfo
  {author} {\bibfnamefont {M.}~\bibnamefont {Salluzzo}}, \bibinfo {author}
  {\bibfnamefont {E.}~\bibnamefont {Schierle}}, \bibinfo {author}
  {\bibfnamefont {R.}~\bibnamefont {Sutarto}}, \bibinfo {author} {\bibfnamefont
  {G.~A.}\ \bibnamefont {Sawatzky}}, \bibinfo {author} {\bibfnamefont
  {E.}~\bibnamefont {Weschke}}, \bibinfo {author} {\bibfnamefont
  {B.}~\bibnamefont {Keimer}}, \ and\ \bibinfo {author} {\bibfnamefont
  {L.}~\bibnamefont {Braicovich}},\ }\Doi {10.1126/science.1223532} {\bibfield
  {journal} {\bibinfo  {journal} {Science},\ }\textbf {\bibinfo {volume}
  {337}},\ \bibinfo {pages} {821} (\bibinfo {year} {2012})}\BibitemShut
  {NoStop}%
\bibitem [{\citenamefont {Achkar}\ \emph {et~al.}(2012)\citenamefont {Achkar},
  \citenamefont {Sutarto}, \citenamefont {Mao}, \citenamefont {He},
  \citenamefont {Frano}, \citenamefont {Blanco-Canosa}, \citenamefont {Tacon},
  \citenamefont {Ghiringhelli}, \citenamefont {Braicovich}, \citenamefont
  {Minola}, \citenamefont {Sala}, \citenamefont {Mazzoli}, \citenamefont
  {Liang}, \citenamefont {Bonn}, \citenamefont {Hardy}, \citenamefont {Keimer},
  \citenamefont {Sawatzky},\ and\ \citenamefont {Hawthorn}}]{Achkar2012}%
  \BibitemOpen
  \bibfield  {author} {\bibinfo {author} {\bibfnamefont {A.~J.}\ \bibnamefont
  {Achkar}}, \bibinfo {author} {\bibfnamefont {R.}~\bibnamefont {Sutarto}},
  \bibinfo {author} {\bibfnamefont {X.}~\bibnamefont {Mao}}, \bibinfo {author}
  {\bibfnamefont {F.}~\bibnamefont {He}}, \bibinfo {author} {\bibfnamefont
  {A.}~\bibnamefont {Frano}}, \bibinfo {author} {\bibfnamefont
  {S.}~\bibnamefont {Blanco-Canosa}}, \bibinfo {author} {\bibfnamefont {M.~L.}\
  \bibnamefont {Tacon}}, \bibinfo {author} {\bibfnamefont {G.}~\bibnamefont
  {Ghiringhelli}}, \bibinfo {author} {\bibfnamefont {L.}~\bibnamefont
  {Braicovich}}, \bibinfo {author} {\bibfnamefont {M.}~\bibnamefont {Minola}},
  \bibinfo {author} {\bibfnamefont {M.~M.}\ \bibnamefont {Sala}}, \bibinfo
  {author} {\bibfnamefont {C.}~\bibnamefont {Mazzoli}}, \bibinfo {author}
  {\bibfnamefont {R.}~\bibnamefont {Liang}}, \bibinfo {author} {\bibfnamefont
  {D.~A.}\ \bibnamefont {Bonn}}, \bibinfo {author} {\bibfnamefont {W.~N.}\
  \bibnamefont {Hardy}}, \bibinfo {author} {\bibfnamefont {B.}~\bibnamefont
  {Keimer}}, \bibinfo {author} {\bibfnamefont {G.~A.}\ \bibnamefont
  {Sawatzky}}, \ and\ \bibinfo {author} {\bibfnamefont {D.~G.}\ \bibnamefont
  {Hawthorn}},\ }\Doi {10.1103/physrevlett.109.167001} {\bibfield  {journal}
  {\bibinfo  {journal} {Phys. Rev. Lett.},\ }\textbf {\bibinfo {volume} {109}}
  (\bibinfo {year} {2012})},\ \doi {10.1103/physrevlett.109.167001}\BibitemShut
  {NoStop}%
\bibitem [{\citenamefont {Chang}\ \emph {et~al.}(2012)\citenamefont {Chang},
  \citenamefont {Blackburn}, \citenamefont {Holmes}, \citenamefont
  {Christensen}, \citenamefont {Larsen}, \citenamefont {Mesot}, \citenamefont
  {Liang}, \citenamefont {Bonn}, \citenamefont {Hardy}, \citenamefont
  {Watenphul}, \citenamefont {v.~Zimmermann}, \citenamefont {Forgan},\ and\
  \citenamefont {Hayden}}]{Chang2012}%
  \BibitemOpen
  \bibfield  {author} {\bibinfo {author} {\bibfnamefont {J.}~\bibnamefont
  {Chang}}, \bibinfo {author} {\bibfnamefont {E.}~\bibnamefont {Blackburn}},
  \bibinfo {author} {\bibfnamefont {A.~T.}\ \bibnamefont {Holmes}}, \bibinfo
  {author} {\bibfnamefont {N.~B.}\ \bibnamefont {Christensen}}, \bibinfo
  {author} {\bibfnamefont {J.}~\bibnamefont {Larsen}}, \bibinfo {author}
  {\bibfnamefont {J.}~\bibnamefont {Mesot}}, \bibinfo {author} {\bibfnamefont
  {R.}~\bibnamefont {Liang}}, \bibinfo {author} {\bibfnamefont {D.~A.}\
  \bibnamefont {Bonn}}, \bibinfo {author} {\bibfnamefont {W.~N.}\ \bibnamefont
  {Hardy}}, \bibinfo {author} {\bibfnamefont {A.}~\bibnamefont {Watenphul}},
  \bibinfo {author} {\bibfnamefont {M.}~\bibnamefont {v.~Zimmermann}}, \bibinfo
  {author} {\bibfnamefont {E.~M.}\ \bibnamefont {Forgan}}, \ and\ \bibinfo
  {author} {\bibfnamefont {S.~M.}\ \bibnamefont {Hayden}},\ }\Doi
  {10.1038/nphys2456} {\bibfield  {journal} {\bibinfo  {journal} {Nat. Phys.},\
  }\textbf {\bibinfo {volume} {8}},\ \bibinfo {pages} {871} (\bibinfo {year}
  {2012})}\BibitemShut {NoStop}%
\bibitem [{\citenamefont {Kharkov}\ and\ \citenamefont
  {Sushkov}(2016)}]{Kharkov2016}%
  \BibitemOpen
  \bibfield  {author} {\bibinfo {author} {\bibfnamefont {Y.~A.}\ \bibnamefont
  {Kharkov}}\ and\ \bibinfo {author} {\bibfnamefont {O.~P.}\ \bibnamefont
  {Sushkov}},\ }\Doi {10.1038/srep34551} {\bibfield  {journal} {\bibinfo
  {journal} {Sci. Rep.},\ }\textbf {\bibinfo {volume} {6}} (\bibinfo {year}
  {2016})},\ \doi {10.1038/srep34551}\BibitemShut {NoStop}%
\bibitem [{\citenamefont {Fradkin}\ \emph {et~al.}(2015)\citenamefont
  {Fradkin}, \citenamefont {Kivelson},\ and\ \citenamefont
  {Tranquada}}]{Fradkin2015}%
  \BibitemOpen
  \bibfield  {author} {\bibinfo {author} {\bibfnamefont {E.}~\bibnamefont
  {Fradkin}}, \bibinfo {author} {\bibfnamefont {S.~A.}\ \bibnamefont
  {Kivelson}}, \ and\ \bibinfo {author} {\bibfnamefont {J.~M.}\ \bibnamefont
  {Tranquada}},\ }\Doi {10.1103/revmodphys.87.457} {\bibfield  {journal}
  {\bibinfo  {journal} {Rev. Mod. Phys.},\ }\textbf {\bibinfo {volume} {87}},\
  \bibinfo {pages} {457} (\bibinfo {year} {2015})}\BibitemShut {NoStop}%
\bibitem [{\citenamefont {Sato}\ \emph {et~al.}(2017)\citenamefont {Sato},
  \citenamefont {Kasahara}, \citenamefont {Murayama}, \citenamefont {Kasahara},
  \citenamefont {Moon}, \citenamefont {Nishizaki}, \citenamefont {Loew},
  \citenamefont {Porras}, \citenamefont {Keimer}, \citenamefont {Shibauchi},\
  and\ \citenamefont {Matsuda}}]{Sato2017}%
  \BibitemOpen
  \bibfield  {author} {\bibinfo {author} {\bibfnamefont {Y.}~\bibnamefont
  {Sato}}, \bibinfo {author} {\bibfnamefont {S.}~\bibnamefont {Kasahara}},
  \bibinfo {author} {\bibfnamefont {H.}~\bibnamefont {Murayama}}, \bibinfo
  {author} {\bibfnamefont {Y.}~\bibnamefont {Kasahara}}, \bibinfo {author}
  {\bibfnamefont {E.-G.}\ \bibnamefont {Moon}}, \bibinfo {author}
  {\bibfnamefont {T.}~\bibnamefont {Nishizaki}}, \bibinfo {author}
  {\bibfnamefont {T.}~\bibnamefont {Loew}}, \bibinfo {author} {\bibfnamefont
  {J.}~\bibnamefont {Porras}}, \bibinfo {author} {\bibfnamefont
  {B.}~\bibnamefont {Keimer}}, \bibinfo {author} {\bibfnamefont
  {T.}~\bibnamefont {Shibauchi}}, \ and\ \bibinfo {author} {\bibfnamefont
  {Y.}~\bibnamefont {Matsuda}},\ }\Doi {10.1038/nphys4205} {\bibfield
  {journal} {\bibinfo  {journal} {Nat. Phys.},\ }\textbf {\bibinfo {volume}
  {13}},\ \bibinfo {pages} {1074} (\bibinfo {year} {2017})}\BibitemShut
  {NoStop}%
\bibitem [{\citenamefont {Bonesteel}\ \emph {et~al.}(1992)\citenamefont
  {Bonesteel}, \citenamefont {Rice},\ and\ \citenamefont
  {Zhang}}]{Bonesteel1992}%
  \BibitemOpen
  \bibfield  {author} {\bibinfo {author} {\bibfnamefont {N.~E.}\ \bibnamefont
  {Bonesteel}}, \bibinfo {author} {\bibfnamefont {T.~M.}\ \bibnamefont {Rice}},
  \ and\ \bibinfo {author} {\bibfnamefont {F.~C.}\ \bibnamefont {Zhang}},\
  }\Doi {10.1103/physrevlett.68.2684} {\bibfield  {journal} {\bibinfo
  {journal} {Phys. Rev. Lett.},\ }\textbf {\bibinfo {volume} {68}},\ \bibinfo
  {pages} {2684} (\bibinfo {year} {1992})}\BibitemShut {NoStop}%
\bibitem [{\citenamefont {Shekhtman}\ \emph {et~al.}(1992)\citenamefont
  {Shekhtman}, \citenamefont {Entin-Wohlman},\ and\ \citenamefont
  {Aharony}}]{Shekhtman1992}%
  \BibitemOpen
  \bibfield  {author} {\bibinfo {author} {\bibfnamefont {L.}~\bibnamefont
  {Shekhtman}}, \bibinfo {author} {\bibfnamefont {O.}~\bibnamefont
  {Entin-Wohlman}}, \ and\ \bibinfo {author} {\bibfnamefont {A.}~\bibnamefont
  {Aharony}},\ }\Doi {10.1103/physrevlett.69.836} {\bibfield  {journal}
  {\bibinfo  {journal} {Phys. Rev. Lett.},\ }\textbf {\bibinfo {volume} {69}},\
  \bibinfo {pages} {836} (\bibinfo {year} {1992})}\BibitemShut {NoStop}%
\bibitem [{\citenamefont {Koshibae}\ \emph {et~al.}(1993)\citenamefont
  {Koshibae}, \citenamefont {Ohta},\ and\ \citenamefont
  {Maekawa}}]{Koshibae1993}%
  \BibitemOpen
  \bibfield  {author} {\bibinfo {author} {\bibfnamefont {W.}~\bibnamefont
  {Koshibae}}, \bibinfo {author} {\bibfnamefont {Y.}~\bibnamefont {Ohta}}, \
  and\ \bibinfo {author} {\bibfnamefont {S.}~\bibnamefont {Maekawa}},\ }\Doi
  {10.1103/physrevb.47.3391} {\bibfield  {journal} {\bibinfo  {journal} {Phys.
  Rev. B},\ }\textbf {\bibinfo {volume} {47}},\ \bibinfo {pages} {3391}
  (\bibinfo {year} {1993})}\BibitemShut {NoStop}%
\bibitem [{\citenamefont {Chakravarty}\ \emph {et~al.}(1988)\citenamefont
  {Chakravarty}, \citenamefont {Halperin},\ and\ \citenamefont
  {Nelson}}]{Chakravarty1988}%
  \BibitemOpen
  \bibfield  {author} {\bibinfo {author} {\bibfnamefont {S.}~\bibnamefont
  {Chakravarty}}, \bibinfo {author} {\bibfnamefont {B.~I.}\ \bibnamefont
  {Halperin}}, \ and\ \bibinfo {author} {\bibfnamefont {D.~R.}\ \bibnamefont
  {Nelson}},\ }\href@noop {} {\bibfield  {journal} {\bibinfo  {journal} {Phys.
  Rev. Lett.},\ }\textbf {\bibinfo {volume} {60}},\ \bibinfo {pages} {1057}
  (\bibinfo {year} {1988})}\BibitemShut {NoStop}%
\bibitem [{\citenamefont {Arovas}\ and\ \citenamefont
  {Auerbach}(1988)}]{Arovas1988}%
  \BibitemOpen
  \bibfield  {author} {\bibinfo {author} {\bibfnamefont {D.~P.}\ \bibnamefont
  {Arovas}}\ and\ \bibinfo {author} {\bibfnamefont {A.}~\bibnamefont
  {Auerbach}},\ }\href@noop {} {\bibfield  {journal} {\bibinfo  {journal}
  {Phys. Rev. B},\ }\textbf {\bibinfo {volume} {38}},\ \bibinfo {pages} {316}
  (\bibinfo {year} {1988})}\BibitemShut {NoStop}%
\bibitem [{\citenamefont {Takahashi}(1989)}]{Takahashi1989}%
  \BibitemOpen
  \bibfield  {author} {\bibinfo {author} {\bibfnamefont {M.}~\bibnamefont
  {Takahashi}},\ }\Doi {10.1103/physrevb.40.2494} {\bibfield  {journal}
  {\bibinfo  {journal} {Phys Rev B},\ }\textbf {\bibinfo {volume} {40}},\
  \bibinfo {pages} {2494} (\bibinfo {year} {1989})}\BibitemShut {NoStop}%
\bibitem [{\citenamefont {Tyablikov}\ and\ \citenamefont
  {Bonch-Bruevich}(1962)}]{Tyablikov1962}%
  \BibitemOpen
  \bibfield  {author} {\bibinfo {author} {\bibfnamefont {S.}~\bibnamefont
  {Tyablikov}}\ and\ \bibinfo {author} {\bibfnamefont {V.}~\bibnamefont
  {Bonch-Bruevich}},\ }\Doi {10.1080/00018736200101312} {\bibfield  {journal}
  {\bibinfo  {journal} {Adv. Phys.},\ }\textbf {\bibinfo {volume} {11}},\
  \bibinfo {pages} {317} (\bibinfo {year} {1962})}\BibitemShut {NoStop}%
\bibitem [{\citenamefont {Kondo}\ and\ \citenamefont
  {Yamaji}(1972)}]{Kondo1972}%
  \BibitemOpen
  \bibfield  {author} {\bibinfo {author} {\bibfnamefont {J.}~\bibnamefont
  {Kondo}}\ and\ \bibinfo {author} {\bibfnamefont {K.}~\bibnamefont {Yamaji}},\
  }\Doi {10.1143/ptp.47.807} {\bibfield  {journal} {\bibinfo  {journal} {Progr.
  Theoret. Phys.},\ }\textbf {\bibinfo {volume} {47}},\ \bibinfo {pages} {807}
  (\bibinfo {year} {1972})}\BibitemShut {NoStop}%
\bibitem [{\citenamefont {Shimahara}\ and\ \citenamefont
  {Takada}(1991)}]{Shimahara1991}%
  \BibitemOpen
  \bibfield  {author} {\bibinfo {author} {\bibfnamefont {H.}~\bibnamefont
  {Shimahara}}\ and\ \bibinfo {author} {\bibfnamefont {S.}~\bibnamefont
  {Takada}},\ }\Doi {10.1143/jpsj.60.2394} {\bibfield  {journal} {\bibinfo
  {journal} {J. Phys. Soc. Jpn.},\ }\textbf {\bibinfo {volume} {60}},\ \bibinfo
  {pages} {2394} (\bibinfo {year} {1991})}\BibitemShut {NoStop}%
\bibitem [{\citenamefont {Winterfeldt}\ and\ \citenamefont
  {Ihle}(1997)}]{Winterfeldt1997}%
  \BibitemOpen
  \bibfield  {author} {\bibinfo {author} {\bibfnamefont {S.}~\bibnamefont
  {Winterfeldt}}\ and\ \bibinfo {author} {\bibfnamefont {D.}~\bibnamefont
  {Ihle}},\ }\Doi {10.1103/physrevb.56.5535} {\bibfield  {journal} {\bibinfo
  {journal} {Phys. Rev. B},\ }\textbf {\bibinfo {volume} {56}},\ \bibinfo
  {pages} {5535} (\bibinfo {year} {1997})}\BibitemShut {NoStop}%
\bibitem [{\citenamefont {Zavidonov}\ and\ \citenamefont
  {Brinkmann}(1998)}]{Zavidonov1998}%
  \BibitemOpen
  \bibfield  {author} {\bibinfo {author} {\bibfnamefont {A.~Y.}\ \bibnamefont
  {Zavidonov}}\ and\ \bibinfo {author} {\bibfnamefont {D.}~\bibnamefont
  {Brinkmann}},\ }\Doi {10.1103/physrevb.58.12486} {\bibfield  {journal}
  {\bibinfo  {journal} {Phys. Rev. B},\ }\textbf {\bibinfo {volume} {58}},\
  \bibinfo {pages} {12486} (\bibinfo {year} {1998})}\BibitemShut {NoStop}%
\bibitem [{\citenamefont {Sadovskii}(2001)}]{Sadovskii2001}%
  \BibitemOpen
  \bibfield  {author} {\bibinfo {author} {\bibfnamefont {M.~V.}\ \bibnamefont
  {Sadovskii}},\ }\Doi {10.1070/pu2001v044n05abeh000902} {\bibfield  {journal}
  {\bibinfo  {journal} {Phys. Usp.},\ }\textbf {\bibinfo {volume} {44}},\
  \bibinfo {pages} {515} (\bibinfo {year} {2001})}\BibitemShut {NoStop}%
\bibitem [{\citenamefont {Bonesteel}(1993)}]{Bonesteel1993}%
  \BibitemOpen
  \bibfield  {author} {\bibinfo {author} {\bibfnamefont {N.~E.}\ \bibnamefont
  {Bonesteel}},\ }\Doi {10.1103/physrevb.47.11302} {\bibfield  {journal}
  {\bibinfo  {journal} {Phys. Rev. B},\ }\textbf {\bibinfo {volume} {47}},\
  \bibinfo {pages} {11302} (\bibinfo {year} {1993})}\BibitemShut {NoStop}%
\bibitem [{\citenamefont {Johnston}(1989)}]{Johnston1989}%
  \BibitemOpen
  \bibfield  {author} {\bibinfo {author} {\bibfnamefont {D.~C.}\ \bibnamefont
  {Johnston}},\ }\Doi {10.1103/physrevlett.62.957} {\bibfield  {journal}
  {\bibinfo  {journal} {Phys. Rev. Lett.},\ }\textbf {\bibinfo {volume} {62}},\
  \bibinfo {pages} {957} (\bibinfo {year} {1989})}\BibitemShut {NoStop}%
\bibitem [{\citenamefont {Nakano}\ \emph {et~al.}(1994)\citenamefont {Nakano},
  \citenamefont {Oda}, \citenamefont {Manabe}, \citenamefont {Momono},
  \citenamefont {Miura},\ and\ \citenamefont {Ido}}]{Nakano1994}%
  \BibitemOpen
  \bibfield  {author} {\bibinfo {author} {\bibfnamefont {T.}~\bibnamefont
  {Nakano}}, \bibinfo {author} {\bibfnamefont {M.}~\bibnamefont {Oda}},
  \bibinfo {author} {\bibfnamefont {C.}~\bibnamefont {Manabe}}, \bibinfo
  {author} {\bibfnamefont {N.}~\bibnamefont {Momono}}, \bibinfo {author}
  {\bibfnamefont {Y.}~\bibnamefont {Miura}}, \ and\ \bibinfo {author}
  {\bibfnamefont {M.}~\bibnamefont {Ido}},\ }\Doi {10.1103/physrevb.49.16000}
  {\bibfield  {journal} {\bibinfo  {journal} {Phys. Rev. B},\ }\textbf
  {\bibinfo {volume} {49}},\ \bibinfo {pages} {16000} (\bibinfo {year}
  {1994})}\BibitemShut {NoStop}%
\bibitem [{\citenamefont {Okabe}\ and\ \citenamefont
  {Kikuchi}(1988)}]{Okabe1988}%
  \BibitemOpen
  \bibfield  {author} {\bibinfo {author} {\bibfnamefont {Y.}~\bibnamefont
  {Okabe}}\ and\ \bibinfo {author} {\bibfnamefont {M.}~\bibnamefont
  {Kikuchi}},\ }\Doi {10.1143/jpsj.57.4351} {\bibfield  {journal} {\bibinfo
  {journal} {J. Phys. Soc. Jpn.},\ }\textbf {\bibinfo {volume} {57}},\ \bibinfo
  {pages} {4351} (\bibinfo {year} {1988})}\BibitemShut {NoStop}%
\bibitem [{\citenamefont {Zhang}\ and\ \citenamefont
  {Rice}(1988)}]{ZhangRice1988}%
  \BibitemOpen
  \bibfield  {author} {\bibinfo {author} {\bibfnamefont {F.~C.}\ \bibnamefont
  {Zhang}}\ and\ \bibinfo {author} {\bibfnamefont {T.~M.}\ \bibnamefont
  {Rice}},\ }\href@noop {} {\bibfield  {journal} {\bibinfo  {journal} {Phys.
  Rev. B},\ }\textbf {\bibinfo {volume} {37}},\ \bibinfo {pages} {3759}
  (\bibinfo {year} {1988})}\BibitemShut {NoStop}%
\bibitem [{\citenamefont {Tohyama}\ and\ \citenamefont
  {Maekawa}(1990)}]{Tohyama1990}%
  \BibitemOpen
  \bibfield  {author} {\bibinfo {author} {\bibfnamefont {T.}~\bibnamefont
  {Tohyama}}\ and\ \bibinfo {author} {\bibfnamefont {S.}~\bibnamefont
  {Maekawa}},\ }\Doi {10.1143/jpsj.59.1760} {\bibfield  {journal} {\bibinfo
  {journal} {J. Phys. Soc. Jpn.},\ }\textbf {\bibinfo {volume} {59}},\ \bibinfo
  {pages} {1760} (\bibinfo {year} {1990})}\BibitemShut {NoStop}%
\bibitem [{\citenamefont {Matsukawa}\ and\ \citenamefont
  {Fukuyama}(1989)}]{Matsukawa1989}%
  \BibitemOpen
  \bibfield  {author} {\bibinfo {author} {\bibfnamefont {H.}~\bibnamefont
  {Matsukawa}}\ and\ \bibinfo {author} {\bibfnamefont {H.}~\bibnamefont
  {Fukuyama}},\ }\Doi {10.1143/jpsj.58.2845} {\bibfield  {journal} {\bibinfo
  {journal} {J. Phys. Soc. Jpn.},\ }\textbf {\bibinfo {volume} {58}},\ \bibinfo
  {pages} {2845} (\bibinfo {year} {1989})}\BibitemShut {NoStop}%
\bibitem [{\citenamefont {Eskes}\ \emph {et~al.}(1989)\citenamefont {Eskes},
  \citenamefont {Sawatzky},\ and\ \citenamefont {Feiner}}]{Eskes1989}%
  \BibitemOpen
  \bibfield  {author} {\bibinfo {author} {\bibfnamefont {H.}~\bibnamefont
  {Eskes}}, \bibinfo {author} {\bibfnamefont {G.}~\bibnamefont {Sawatzky}}, \
  and\ \bibinfo {author} {\bibfnamefont {L.}~\bibnamefont {Feiner}},\ }\Doi
  {10.1016/0921-4534(89)90415-2} {\bibfield  {journal} {\bibinfo  {journal}
  {Physica C},\ }\textbf {\bibinfo {volume} {160}},\ \bibinfo {pages} {424}
  (\bibinfo {year} {1989})}\BibitemShut {NoStop}%
\bibitem [{\citenamefont {Hybertsen}\ \emph {et~al.}(1990)\citenamefont
  {Hybertsen}, \citenamefont {Stechel}, \citenamefont {Schluter},\ and\
  \citenamefont {Jennison}}]{Hybertsen1990}%
  \BibitemOpen
  \bibfield  {author} {\bibinfo {author} {\bibfnamefont {M.~S.}\ \bibnamefont
  {Hybertsen}}, \bibinfo {author} {\bibfnamefont {E.~B.}\ \bibnamefont
  {Stechel}}, \bibinfo {author} {\bibfnamefont {M.}~\bibnamefont {Schluter}}, \
  and\ \bibinfo {author} {\bibfnamefont {D.~R.}\ \bibnamefont {Jennison}},\
  }\Doi {10.1103/physrevb.41.11068} {\bibfield  {journal} {\bibinfo  {journal}
  {Phys. Rev. B},\ }\textbf {\bibinfo {volume} {41}},\ \bibinfo {pages} {11068}
  (\bibinfo {year} {1990})}\BibitemShut {NoStop}%
\end{thebibliography}%

\end{document}